\documentstyle{report}

\begin{document}

WE ARE NOT STUCK WITH GLUING a response to a note of A. Ocneanu

\smallskip

by D. Yetter and L. Crane

\smallskip

In [1], we outlined a procedure for constructing a 4D topological
Quantum Field Theory(TQFT) from a modular tensor category (MTC).

The construction is related to the well known construction of a
3d tqft . In our announcement we gave the formula for the invariant
as follows:
\bigskip

\[ \sum \hspace{1ex}
N^{\# vertices - \# edges} \prod_{faces} dim_q(j)
\prod_{tetrahedra} dim_q^{-1}(p) \prod_{4-simplexes} 15J_q \hspace{.2in}(*) \]
\bigskip

\noindent where the sum ranges over all assignments of spins to the
faces and tetrahedra of the triangulation and $j$ represents the spin
labelling a face, $p$ represents the spin labelling the cut interior
to a tetrahedron, $dim_q$ is the quantum dimension, and $ N$ is
the sum of the squares of the quantum dimensions. Here by spins, we
mean irreducible representations of quantized $sl_2$ at a root of
unity.

\smallskip

We have two different ways of thinking of our quantum 15J symbols.
One, which really plays a heuristic role for us, is as an invariant of
a labelled surface embedded in $S^3$. The other, which we use directly in
our proof, is as a recombination diagram in a braided tensor category.
Perhaps we have been a little too cavalier in using the first picture,
since the connection between the two involves some subtleties.

In [2], A. Ocneanu announced the result that the invariant we define
is always 1, and asserted that our procedure is equivalent to one he
examined earlier, in a different context.

Although we think that professor Ocneanu's argument is interesting,
and in fact that the construction he suggests is of interest even if
it does give 1 for any closed 4 manifold, we do not believe that the
two constructions are the same. In particular, we know by direct
calculation that our invariant is not constant, nor is it 1 for all simple
cases.

The point of departure in [2] is the assertion that the formula above
is the same as gluing of the 3 manifolds with boundary which are
related to the 15J-q symbols defined in [1]. We do not see how this
could be the case. Note that in our formula the internal and external
spins do not enter in the same way. Gluing would be regarding the 15J
symbol as coming from a manifold with boundary, in which the external
and internal spins play identical roles. Thus our formula does not
appear to have the proper symmetry to express gluing.

There seems to be no way to make professor Ocneanu's results agree
with our calculations. In the first place, he does not get a result
which is independant of the triangulation of the 4 manifold.
The formula he computed reduces the invariant of a 4 manifold to
one for a connected sum of copies of $S^3 \times S^1$ [in a 3D TQFT], where the
number of copies depends on the triangulation chosen.

If we are to take it that our formula, as normalized, is equivalent to
gluing, then we are being told that a topological invariant is equal
to a non-invariant. If we are to believe that our formula without the
normalization is equal to gluing, we run into the immediate problem
that when we join 3 15J symbols around a common face then we do not
get the result corresponding to gluing topologically, but rather an
extra factor corresponding to a loop which is split off, which
corresponds to our factors of N.

If our process were in fact gluing, then the same arguments which
allow us to join together parts of the boundary surfaces corresponding
to disjoint 15J's would also allow us to join separate segments of a
connected boundary surface to itself. In fact, the combination rules
in the category which allow us to join disjoint categorical diagrams
do not extend to that case. Indeed, before realizing this, we briefly
thought that our invariant would be quite simple [although certainly
not constant].

Furthermore, if we take it that the invariant of a connected sum of
$S^1 \times S^1$'s is always 1, we would be led to the conclusion that our
invariant was always 1. ( Probably any constant could be absorbed as a
normalization). This, however, contradicts the calculations we have
been able to do by hand.

Our calculations show that the invariant of $S^4$ is N (= 2 for r=3,
=4 for r=4) . Calculating the invariant of
$S^3 \times S^1$ is more complicated , but yields 1 for r=3,4, not
agreeing with the number for $S^4$. The case of $S^2 \times S^2$ is much
more complicated. Our initial calculuations, which we have not
thoroughly
checked yield an expression involving the braiding.

We are left with the problem of how often we obtain a power of
N as invariant, and whether some modification of Ocneanu's argument
could tell us that.

The interest in the theory we construct does not reduce to the
invariants of compact 4 manifolds. In fact for possible applications
to quantum gravity, compact 4 manifolds are irrelevant, since compact
spacetimes are not causal. It is also possible that the relative form
of our construction for manifolds with boundary could give invariants
of embedded surfaces which are richer than the invariants of closed 4
manifolds. For these reasons, we think that Ocneanu's construction may
also be of considerable interest, regardless of its triviality on
closed 4 manifolds.

Finally, we still do not know if our invariants can distinguish
homeomorphic 4 manifolds. We can see no solution to this problem,
except to compute them for some examples like Dolgachev surfaces.

In summary, we believe that professor Ocneanu's assumption that our
formula is equivalent to gluing, although a natural hypothesis, is not
supported by the facts.

\bigskip

REFERENCES:
\smallskip

1.L. Crane and D. Yetter A Categorical Construction of 4D Topologocal
Quantum Field Theories ksu preprint

\smallskip

2.A. Ocneanu A Note on Simplicial Dimension Shifting preprint

\end{document}